\documentclass[prb,twocolumn,showpacs,preprintnumbers,amsmath,amssymb]{revtex4-1}

\usepackage{graphicx}
\usepackage{dcolumn}
\usepackage{bm}

\begin{document}

\title{Itinerant origin of the ferromagnetic quantum critical point
in Fe(Ga,Ge)$_3$}

\author{David J. Singh}

\affiliation{Materials Science and Technology Division,
Oak Ridge National Laboratory, Oak Ridge, Tennessee 37831-6056}

\date{\today}

\begin{abstract}
The electronic structure and magnetic properties of FeGa$_3$ and
doped FeGa$_3$ are studied using density functional calculations.
An itinerant mechanism for ferromagnetism is found both for $n$-type
doping with Ge and also for $p$-type doping. Boltzmann transport
calculations of the thermopower are also reported.
\end{abstract}

\pacs{}

\maketitle

\section{Introduction}

FeGa$_3$ is a tetragonal
Fe containing semiconductor with a band gap of $\sim$0.5 eV,
\cite{haussermann} and interesting thermoelectric properties.
These include a high thermopower
when doped, \cite{amagai,hadano,hald} although high figures
of merit $ZT$ have not been realized due to the combination of
thermal and electrical conductivity.
The compound shows non-magnetic behavior, but remarkably
when modestly
electron doped by Ge a ferromagnetic quantum critical point
emerges and the ground state becomes a ferromagnetic
metal. \cite{umeo,hald2}
Such quantum
critical systems can show unconventional and sometimes remarkable
physical properties, especially if the magnetic system is strongly
coupled to itinerant electrons and if the itinerant electrons
are heavy. From an experimental point of view, FeGa$_3$ does show
evidence that it is in this regime, from transport, specific
heat and other measurements. \cite{umeo,bittar,storchak}
For example, Umeo and co-workers report a specific heat $\gamma$
of 70 mJ K$^{-2}$ per mole for FeGa$_{3-y}$Ge$_y$, $y$=0.09,
while the thermopower $S(T)$, as mentioned is large, e.g.
$S(300)\sim$-400 -- -560 $\mu$V/K.
for $n$-type carrier concentrations in the 10$^{18}$ cm$^{-3}$ range.
\cite{amagai,hadano,hald}

Several density functional theory (DFT) studies using standard
functionals (with no additional correlation term, such as $U$ in
LDA+$U$ calculations) have shown that the band gap of FeGa$_3$
is well described without magnetism on the Fe and without strong
correlation effects.
\cite{haussermann,imai,yin,guillen,verchenko}
However, Yin and Pickett also reported calculations with an additional
interaction $U$ in LDA+$U$ calculations. 
\cite{yin}
They observed that at
modest values of $U-J$=1.4 eV ($J$ is the Hund's parameter in the LDA+$U$
calculations, the dependence is on $U-J$) an antiferromagnetic
state appears, with moments somewhat below 1 $\mu_B$, which is the spin 1/2
value.
It has subsequently been argued that the observed magnetism
in doped FeGa$_3$ is related to these moments, specifically
that the additional carriers break up singlets formed by spin 1/2 Fe dimers
in the structure leading to free spins that order ferromagnetically. \cite{hald2}

There are some questions related this explanation.
First of all the undoped compound is diamagnetic
below room temperature,
\cite{haussermann,tsujii,hadano} and
the susceptibility shows an increase above
room temperature consistent with free carriers generated for
a band gap in accord with the measured band gap. \cite{tsujii}
Secondly, $^{57}$Fe Mossbauer spectra show no magnetism in the
undoped compound.
Third, the thermoelectric transport properties of RuGa$_3$ are
OsGa$_3$ indicate similar behavior to FeGa$_3$ allowing for
carrier concentration differences; \cite{amagai,takagiwa}
this indicates a similar electronic structure.
Finally, while Yin and Pickett \cite{yin} indicate that they obtain
qualitatively similar behavior and predictions independent
of the +$U$ double counting scheme (i.e. fully localized
limit (FLL or SIC))
or around mean field (AMF)), Osorio-Guillen and co-workers \cite{guillen}
report that they find no local
moments with an optimized double counting scheme
intermediate between these limits.

Here we present first principles calculations showing that the
magnetism of doped FeGa$_3$ can be readily explained in an itinerant
picture without the need for pre-existing moments in the semiconducting
state and without the need for correlation terms. We also present
Boltzmann transport calculations of the thermopower and
a resolution of the differences between the results of Yin and Pickett
\cite{yin} and those of Osorio-Guillen and co-workers. \cite{guillen}

\section{Computational Details}

The present calculations were performed within density functional
theory using the generalized gradient approximation of Perdew,
Burke and Ernzerhof (PBE). \cite{pbe}
For this purpose we used the general potential linearized
augmented planewave (LAPW) method \cite{singh-book}
as implemented in the WIEN2k code. \cite{wien2k}
We used LAPW sphere radii of 2.05 Bohr for both Fe and Ga,
with highly converged basis sets.
We used the standard LAPW basis set with additional local orbitals
rather than the more efficient but potentially less
accurate APW+lo method. We did calculations both in a scalar
relativistic approximation and also including spin orbit.
We did not find significant differences between these.
We also did GGA+$U$ calculations both with the fully
localized limit (SIC) double counting and the around mean field (AMF)
double counting. For consistency, the GGA+$U$ results
discussed here were also done with the PBE GGA.

We used the experimental crystal structure, i.e.
tetragonal $P4_2/mnm$, $a$=6.2628 \AA, $c$=6.5546 \AA,
Fe at (0.3437,0.3437,0), Ga1 at (0,0.5,0), Ga2 at (0.1556,0.1556,0.262),
and four formula units per unit cell.
Each Fe atom in this structure has eight Ga neighbors at distances
between 2.36 \AA{} and 2.50 \AA, which is an arrangement that is
not favorable for strong $d$ bonding, as well as one nearby Fe atom
at $\sim$2.77 \AA. These relatively short bonded Fe-Fe pairs are
the dimers discussed by Yin and Pickett. \cite{yin}
Interestingly, the calculated forces on the atoms were zero to the
precision of the calculation
(largest force was below 2 mRy/Bohr in
scalar relativistic calculations) for this structure. This is in contrast to the
Fe-based superconductors, where non-magnetic density functional
calculations yield large errors in the structure relative to
experiment, a fact that is thought
to be related to the interplay
between bonding and magnetism in those compounds.
\cite{mazin-mag,bondino}

\section{electronic structure}

\begin{figure}
\includegraphics[width=\columnwidth,angle=0]{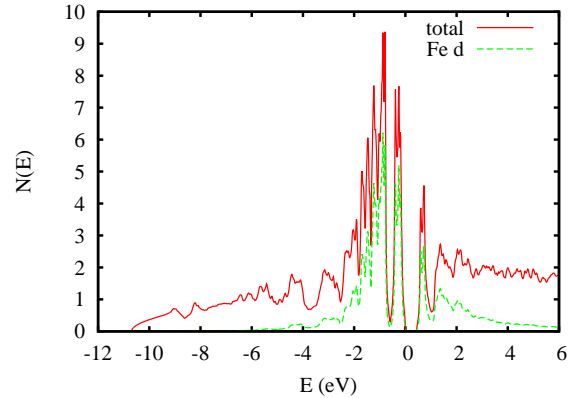}
\includegraphics[width=\columnwidth,angle=0]{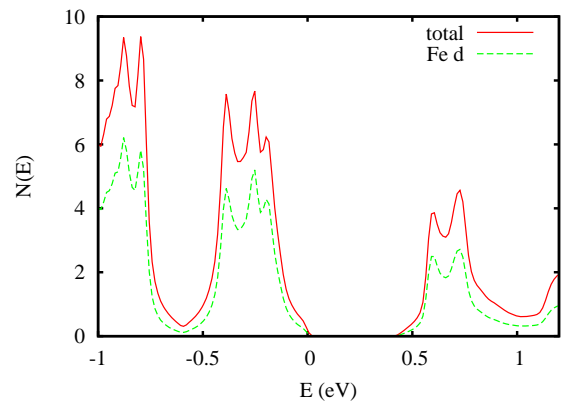}
\caption{(color online) Calculated electronic density of states and Fe $d$
projection onto the Fe LAPW sphere for
FeGa$_3$ on a per formula unit basis. Spin orbit is included.
The bottom panel is a blow-up around the band gap.}
\label{dos}
\end{figure}

The electronic density of states (DOS) as obtained with the PBE GGA
is shown in Fig. \ref{dos}.
The valence electronic structure
is semiconducting and consists of a broad set of Ga derived $sp$
bands starting at $\sim$-11 eV with respect to the valence band
maximum (VBM). A much narrower set of Fe $d$ bands overlap these
and extend from $\sim$-2 eV to $\sim$+2 eV with respect to the VBM.
The Fe $d$ derived part of the DOS (Fig. \ref{dos}) consists of narrow
peaks and makes practically no contribution below $\sim$-3 eV binding energy
and certainly not near the bottom of the Ga $sp$ bands, indicating relatively
weak bonding between the Fe $d$ and Ga $sp$ systems.

\begin{figure}
\includegraphics[width=\columnwidth,angle=0]{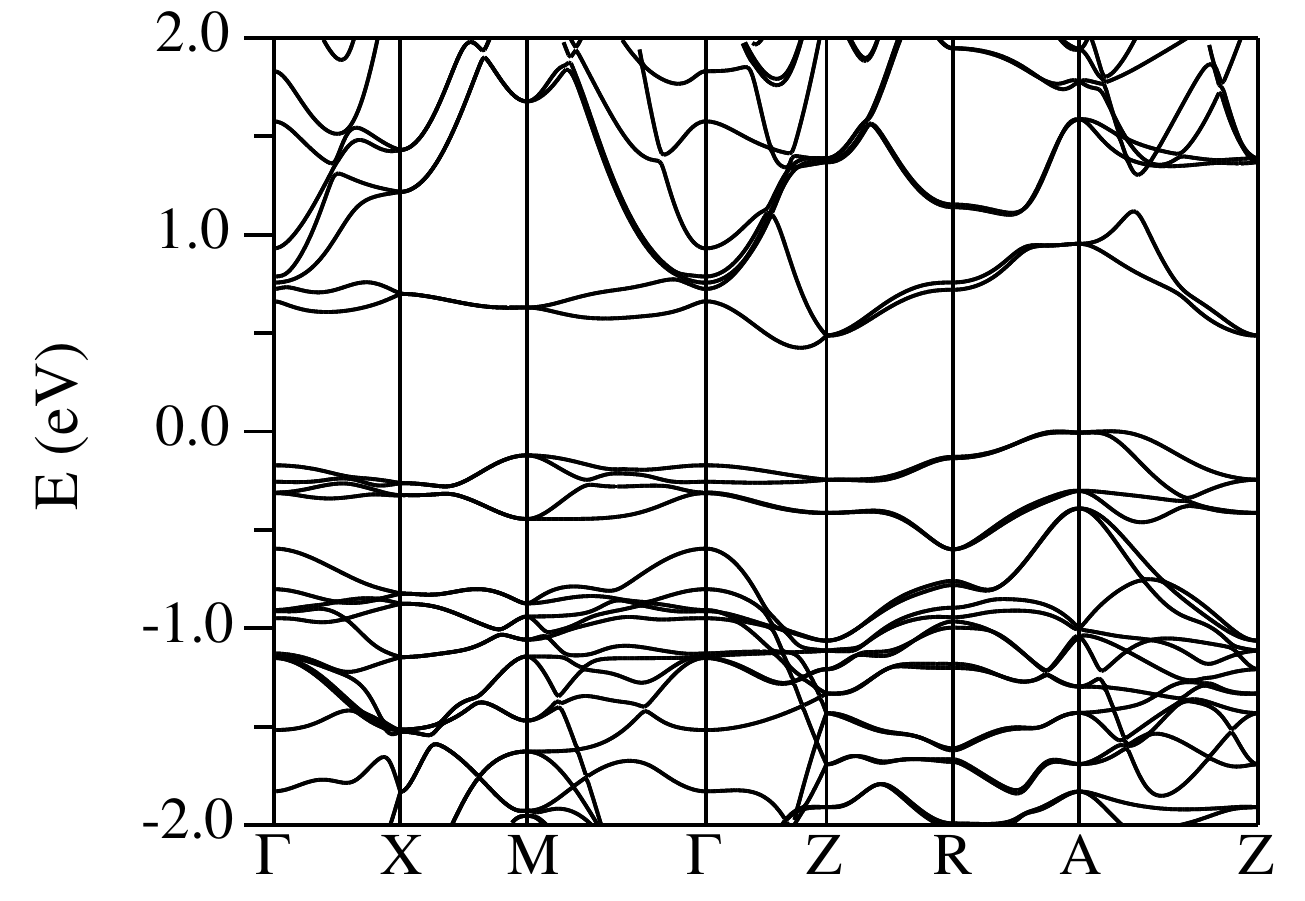}
\caption{Band structure of FeGa$_3$ including spin orbit as
obtained with the PBE GGA.}
\label{bands}
\end{figure}

The band structure in the energy
region around the band edge is shown in Fig. \ref{bands}.
The calculated band gap is $E_g$=0.426 eV including spin orbit,
and becomes only slightly larger (0.428 eV) if spin orbit is neglected.
This small difference presumably
reflects the fact that the states at the band edges come from
very narrow ($\leq$0.5 eV wide)
Fe $d$ bands with little Ga $p$ contribution (spin orbit
is stronger for $p$ states than $d$ states and for heavier atoms,
i.e. Ga rather than Fe). It is also
notable that the valence and
conduction bands have different orbital characters.
With the crystallographic
setting above the conduction band minimum (CBM) has primarily
(the low site symmetry leads to mixing of the $d$-orbitals) $z^2$ orbital
character, while the VBM has primarily $x^2-y^2$ character.

\begin{figure}
\includegraphics[width=\columnwidth,angle=0]{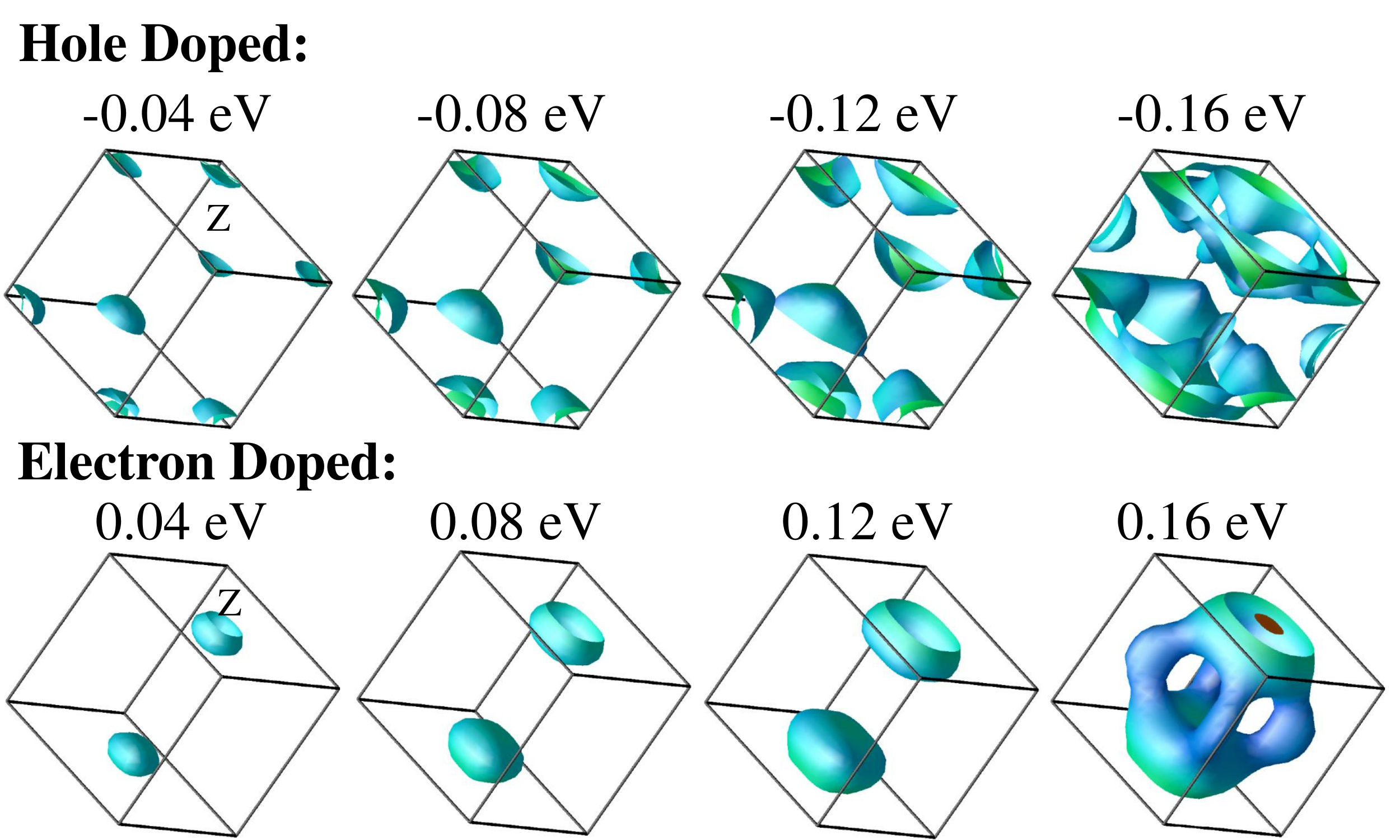}
\caption{(color online)
Rigid band Fermi surface for shifts of $E_F$ into the
valence (top) or conduction (bottom) bands. The energies
are $E_F$ relative to the VBM or CBM, respectively.
The shading is by velocity.
Spin orbit is included.
The corresponding hole or electron counts in parenthesis
on a per unit cell, four formula unit basis, are for hole doped
-0.16 eV (0.93), -0.12 eV (0.43), -0.08 eV (0.21), -0.04 eV (0.09) and
for electron doped, 0.04 eV (0.03), 0.08 eV (0.08), 0.12 eV (0.17),
and 0.16 eV (0.51).
}
\label{fermi-2}
\end{figure}

Turning to the structure of the DOS, one notes that the
DOS increases very rapidly away from the band edges reflecting
the narrow bands and reaches high values well above 2 eV$^{-1}$ on
a per Fe basis. This suggests the possibility of a Stoner mechanism 
for itinerant ferromagnetism when doped. Such a steep density of
states in a material that can be doped metallic is also favorable
for obtaining high thermopowers at high doping levels. This is
one ingredient in obtaining thermoelectric performance (the others
are low lattice thermal conductivity and high mobility;
the figure of merit is $ZT=\sigma S^2T/\kappa$, where $\sigma$ is
the electrical conductivity, $\kappa$ is the thermal
conductivity and $S$ is the thermopower).

\section{Fermi surfaces and transport}

Before discussing the magnetism, we briefly mention the
Fermi surface for doped material and the thermopower.
Fig. \ref{fermi-2} shows the calculated Fermi surface for rigid
band shifts of the Fermi energy, $E_F$ into the conduction and
valence band edges. As may be seen, besides having different
orbital character, the Fermi surfaces for hole and electron doped
material are very different. For lightly electron doped material,
the Fermi surface consists of a three dimensional electron section
near the $Z$ point. At low carrier concentration this is a pocket
off the $Z$ point along the $\Gamma$-$Z$ direction corresponding to
the CBM in Fig. \ref{bands}. The two pockets on the opposite sides
of the $k_z$=1/2 zone boundary then merge to form the $Z$ point pocket,
which has a narrowing in the $k_z$=1/2 plane as seen. The critical
composition at which ferromagnetism is reported to
start is at $x_c$=0.043, \cite{umeo}
which corresponds to a band filling between the 0.08 and 0.12 eV plots
of Fig. \ref{fermi-2}.

As the carrier density further increases, a second
pocket develops around $Z$ and the Fermi surface forms connections
along the $k_z$ direction, as shown in the bottom right panel of
the figure.
For higher carrier concentration (not shown) these
connections merge to form a complex shaped cylinder while $Z$
centered sections remain.
Besides observing the complexity of the Fermi surface, one may note
that it is clearly three dimensional with substantial dispersion in
both the in-plane and $k_z$ directions. This is in contrast to
other well studied
layered materials that may be near ferromagnetic quantum critical
points, i.e. Na$_x$CoO$_2$, \cite{singh-nac,singh-nacf,helme,bayrakci,cao}
and Sr$_3$Ru$_2$O$_7$. \cite{grigera}
The Fermi surfaces for hole doping are pockets around the $A$ point.
These connect in the $k_z$=0.5 plane for higher doping levels.
As seen these are also very three dimensional.

We calculated the doping dependent thermopower with a rigid band
approximation and the constant scattering time approximation, similar
to recent studies on thermoelectric materials.
\cite{parker-bi2se3,singh-pbte,chen-cu2o}
This was done using the BoltzTraP code, \cite{boltztrap}
which employs a smooth interpolation of the energy bands on a fine
grid to calculate band velocities and perform transport integrals.
Importantly, the constant scattering time approximation allows
one to obtain first principles results for the thermopower $S(T)$
without any adjustable parameters.

\begin{figure}
\includegraphics[width=0.9\columnwidth,angle=0]{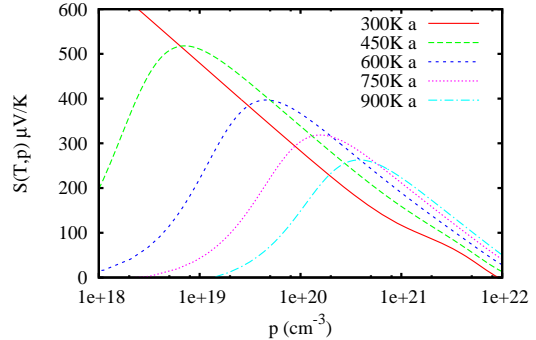}
\includegraphics[width=0.9\columnwidth,angle=0]{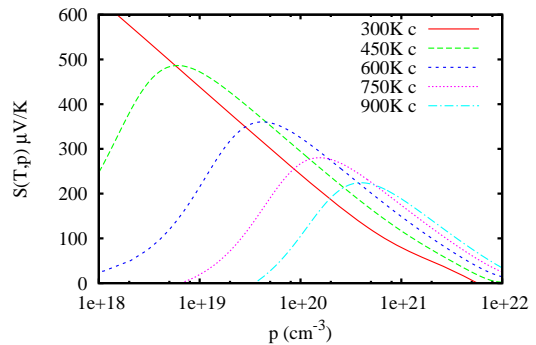}
\caption{(color online) Calculated $p$-type $S(T)$, for in-plane
(top) and $c$-axis (bottom) transport, as a function of carrier
concentration.}
\label{seeb-p}
\end{figure}

\begin{figure}
\includegraphics[width=0.9\columnwidth,angle=0]{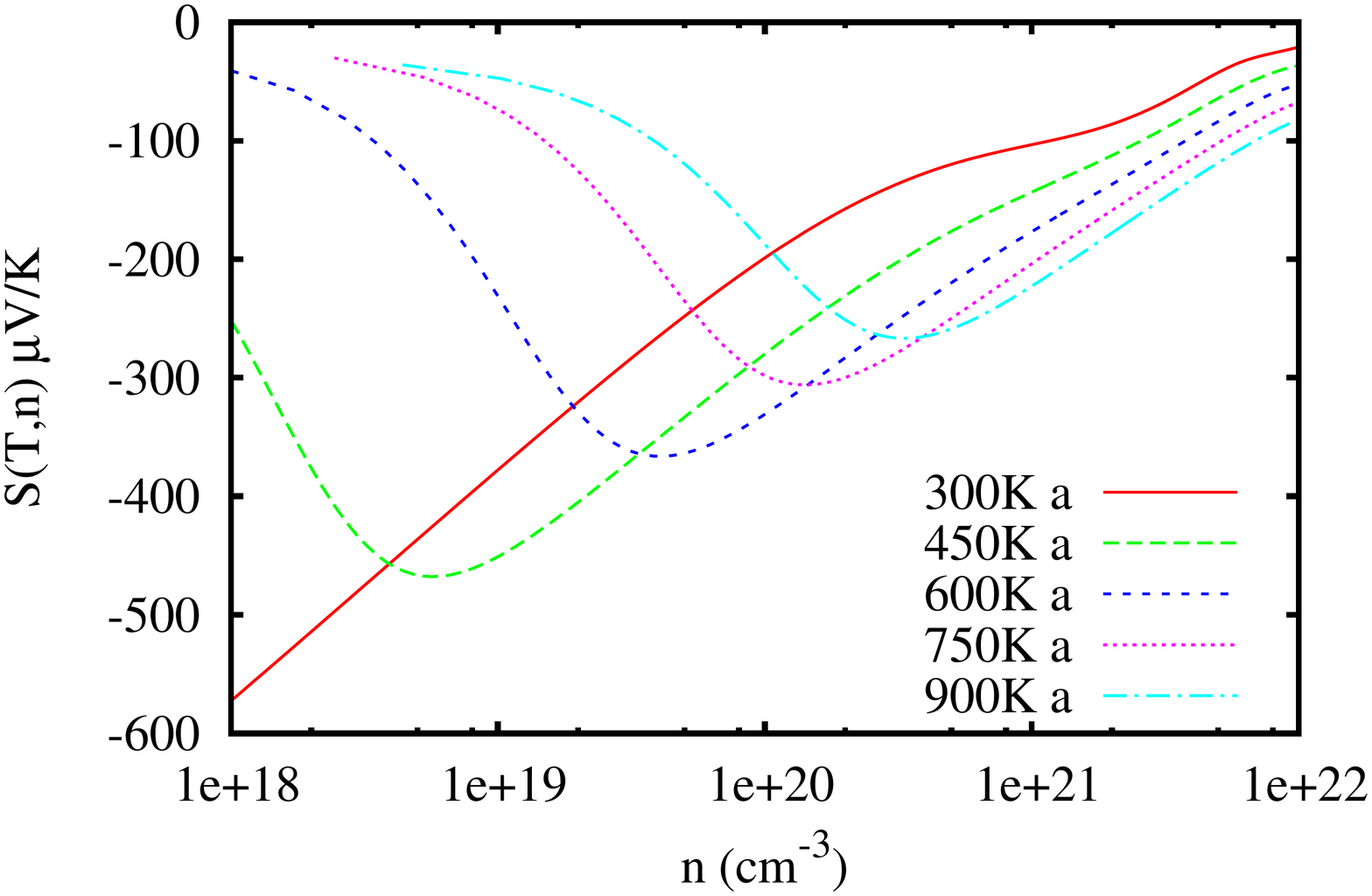}
\includegraphics[width=0.9\columnwidth,angle=0]{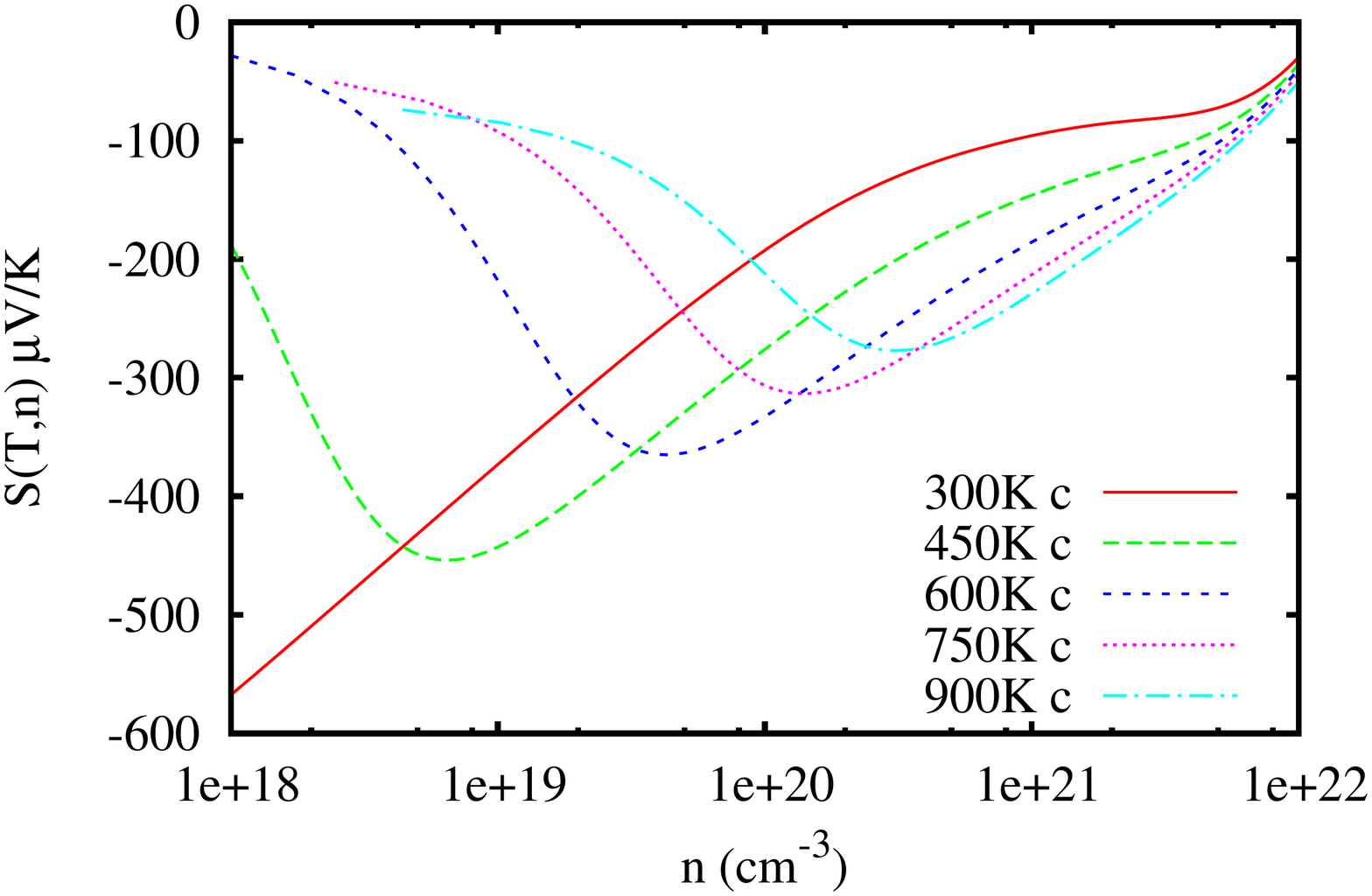}
\caption{(color online) Calculated $n$-type $S(T)$, for in-plane
(top) and $c$-axis (bottom) transport, as a function of carrier
concentration.}
\label{seeb-n}
\end{figure}

The calculated $S(T)$ for $p$-type and $n$-type FeGa$_3$ are shown
in Figs. \ref{seeb-p} and \ref{seeb-n}, respectively.
As may be seen, the behavior is rather symmetric between $p$ and $n$
type in that high values of $S(T)$ are obtained at relatively high
doping levels $\sim$10$^{20}$ cm$^{-3}$. Also the thermopower
is rather isotropic as is typically but not always the case. \cite{ong}

Hadano and co-workers \cite{hadano} obtained an
$n$-type thermopower reaching $\sim$-350 $\mu$V/K at 300 K, which is
consistent with the highest thermopower at that temperature in our
calculations. The inferred carrier concentration is then
$n\sim 10^{19}$ cm$^{-3}$ or slightly
below, and the downturn at higher $T$ is
presumably due to bi-polar transport.
Amagai and co-workers
\cite{amagai} obtained $S(300 K)$=-563 $\mu$V/K on a sample
with a Hall carrier concentration of 3.1x10$^{18}$ cm$^{-3}$.
Taking the Hall concentration as the absolute carrier concentration,
we obtain $S(300 K)$=-480 $\mu$V/K for this condition.
Haldolaarachchige and co-workers \cite{hald}
also reported high
room temperature values, though somewhat lower than those of Amagai 
and co-workers, on nominally stoichiometric FeGa$_3$, with a decrease
upon doping with Co and Ge.

The magnitudes of $S(T)$ are relatively high compared to
most semiconductors, reflecting the narrow bands around the band edges.
These are sufficient that one might expect good thermoelectric performance
in this compound if the doping level
can be optimized and the other properties are favorable.
In this regard the reported lattice
thermal conductivity of FeGa$_3$ is 3.7
W/mK at 300 K and decreases to $\sim$ 2 W/mK at at 900 K. \cite{amagai}
These are reasonably low values consistent with a thermoelectric
material that can be used at high $T$. We note that
the $T$ dependence suggests that some of the reported
lattice thermal conductivity at high $T$ could in fact
have an origin in the bi-polar conduction.
In any case, the thermal conductivity was observed to decrease with
doping. \cite{hald}
While the values of the experimental thermoelectric figures of merit reported
to date are not high, the present results for the thermopower
suggest that the best performance would be obtained at high $T$
(750 K and above) with carrier concentrations of 3x10$^{20}$ cm$^{-3}$
- 1x10$^{21}$ cm$^{-3}$ for both $p$-type and $n$-type.
This is a regime for which thermoelectric
properties have not been reported. It will be
of interest to study samples in this range of carrier concentration
and temperature.

\section{doping and ferromagnetism}

\begin{figure}
\includegraphics[width=\columnwidth,angle=0]{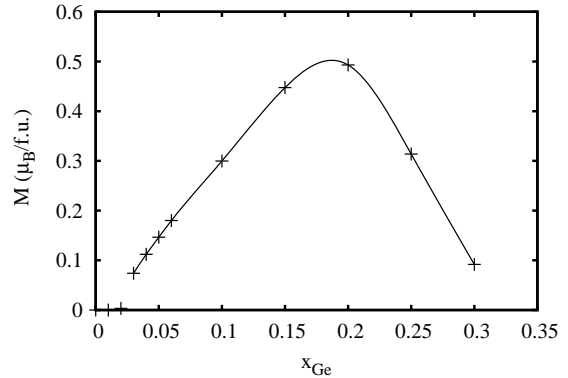}
\caption{Total magnetic moment per formula unit as a function of
composition for Fe(Ga$_{1-x}$Ge$_x$)$_3$ as obtained with the
virtual crystal approximation including spin orbit. This figure is based on
the total magnetization integrated over the whole unit cell. The moments
inside the Fe LAPW spheres (radius 2.05 Bohr) account for
$\sim$75\% of the magnetization}.
\label{mag}
\end{figure}

\begin{figure}
\includegraphics[width=\columnwidth,angle=0]{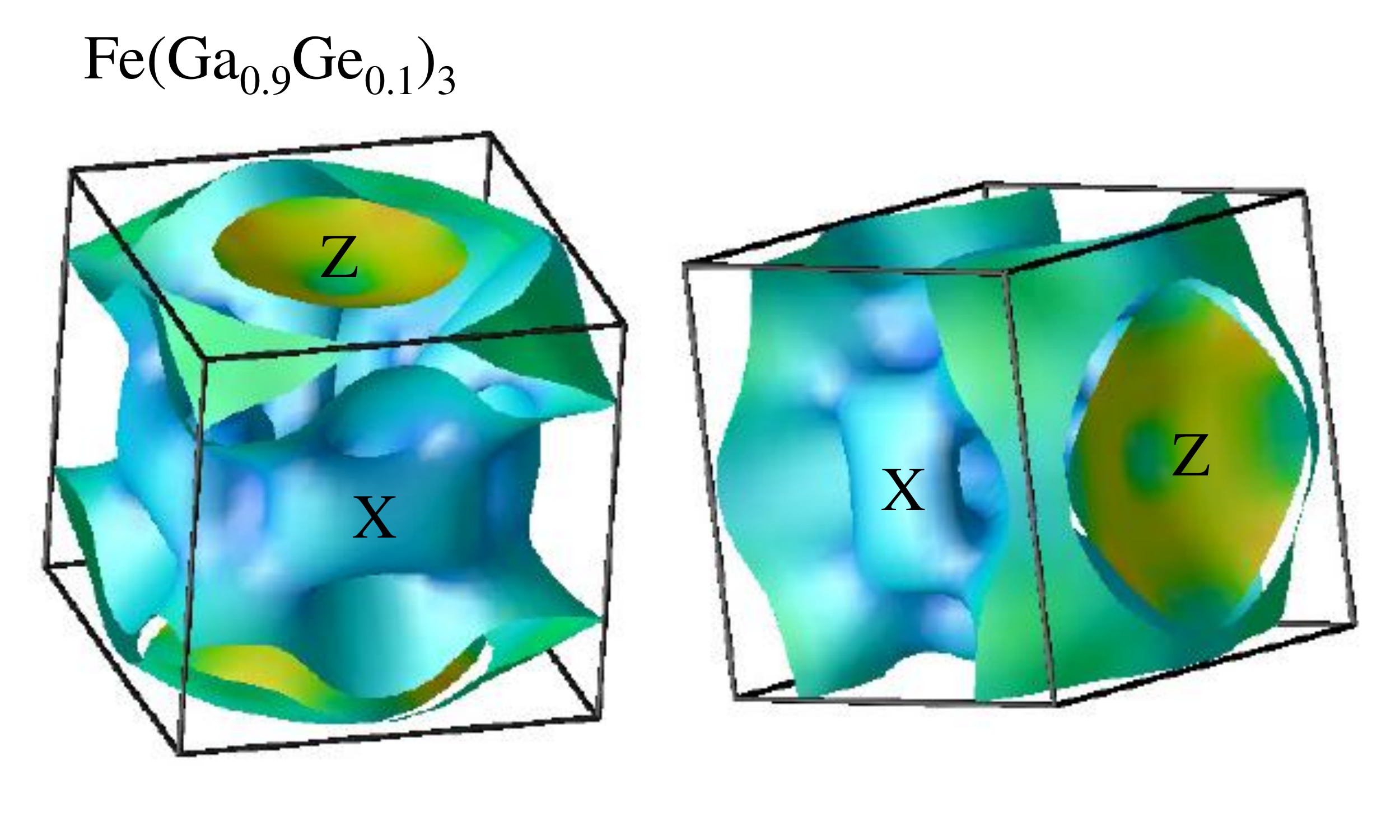}
\caption{(color online)
Calculated Fermi surface of virtual crystal ferromagnetic
Fe(Ga$_{0.9}$Ge$_{0.1}$)$_3$ as obtained with the PBE GGA, including
spin-orbit. The shading is by velocity.}
\label{fermi}
\end{figure}

We examined the possibility of itinerant
Stoner ferromagnetism using direct calculations.
These included virtual crystal calculations, with the virtual
crystal approximation applied to the Ga site as well as
supercell calculations in which a Ge atom was substituted
on either the Ga1 or Ga2 sites in cell
(i.e. $x_{Ge}$=1/12). We also did calculations
where one of the four Fe atoms in the unit cell was replaced by Co.
In these calculations with partial substitution, we kept the lattice
parameters fixed, but relaxed the atomic coordinates by total energy
minimization. These three ordered cells all contain one electron
per unit cell more than stoichiometric FeGa$_3$.

We start with the virtual crystal approximation, which allows for
arbitrary doping levels.
The main result is shown in Fig. \ref{mag}.
We find itinerant ferromagnetism starting at low doping levels,
$\sim x_{Ge}=$0.03. This becomes half metallic almost immediately
as the doping level is increased. This half metallic
ferromagnetic state persists up to $\sim x_{Ge}=$0.15, after
which the moment saturates and turns down above $\sim x_{Ge}=$0.2,
i.e. going over to an ordinary ferromagnetic metal. The ferromagnetism
is lost near $x_{Ge}$=0.3

Half metals are ferromagnets
where one spin channel is semiconducting, while the other is metallic.
\cite{degroot,coey,muller}
In such materials, spin flip scattering is blocked and
ferromagnetic domain walls can have high resistance,
leading to unusual magnetotransport phenomena such as large negative
magnetoresistance.
Such materials are also of interest as ``spin-tronic" materials,
since their electrical transport is entirely in one spin channel,
and also because the magnetic excitation spectrum is different
than in ordinary metallic ferromagnets, in particular as regards
the Stoner continuum.
In the present case, the majority spin channel is metallic
while the minority spin is semiconducting.
It is interesting to note that the calculated behavior
regime is similar to that of the $p$-type
material, Na$_x$CoO$_2$. \cite{singh-nacf}
Na$_x$CoO$_2$ is
also a good thermoelectric, whose high thermopower can be associated
with narrow transition metal $d$ bands. \cite{xiang}

In any case,
the behavior found is consistent with experimental reports. \cite{umeo}
The critical Ge concentration in experiment is reported to be 
$x_c=$0.043, which is close to but
above the calculated value. This is in contrast
to standard density functional results for other quantum critical
systems. In several of
those cases the magnetic state is overly stable relative to experiment,
reflecting neglect of quantum critical fluctuations in standard
density functional calculations. \cite{moriya-book,singh-nacf}
This may imply that the magnitude of the critical fluctuations in
Fe(Ga$_{1-x}$Ge$_x$)$_3$ are weaker than those in e.g. Sr$_3$Ru$_2$O$_7$,
which would imply a smaller region in temperature-composition
(corresponding to temperature-field in Sr$_3$Ru$_2$O$_7$)
space showing quantum criticallity
associated with nearness to ferromagnetism.

The calculated Fermi surface for ferromagnetic Fe(Ga$_{0.9}$Ge$_{0.1}$)$_3$ is
shown in Fig. \ref{fermi}. The material is half-metallic at this composition.
The composition
corresponds to a charge of 1.2 additional electrons per unit cell
relative to FeGa$_3$. Not surprisingly, the Fermi surface is large
and complex for this high electron count. As mentioned, it is
clearly not a two dimensional electron system. This illustrates the continuous
evolution of the material with doping from a low carrier density ferromagnetic
semiconductor to a half metallic large Fermi surface metal.

It is not known how to heavily dope $p$-type in this compound.
However, not surprisingly in view of the DOS, we also find in virtual
crystal calculations that
the compound will become a half-metallic
ferromagnetic metal in this case. It
will be of interest to investigate potential $p$-type dopants to
determine whether a ferromagnetic quantum critical point can be
obtained in this case as well.

As mentioned, calculations were also done for ordered cells. We find
half metallic ferromagnetism very similar to that in the virtual
crystal case when one of the twelve Ga in the unit cell is replaced
by Ge. The density of states is practically the same as in the
virtual crystal calculation for this case. In the cell where one Fe
was replaced by Co, we also find a density of states near the band edges
that is similar to the undoped case and we find an upward shift of the
Fermi energy into the conduction band. We find therefore that Co is an effective
$n$-type dopant similar to Ge and that
the Co-doped system shows near rigid band
behavior. This is similar to what was found
by Haussermann and co-workers for pure CoGa$_3$ in
relation for FeGa$_3$, \cite{haussermann}
and is also in accord with the calculations of Verchenko and co-workers.
\cite{verchenko}
Consistent with this we obtain a half
metallic ferromagnetic state in this cell. This is in apparent disagreement
with experiment, where it is reported that Co substitution does not 
produce ferromagnetism in any amount. \cite{umeo} We do not know
the reason for this disagreement. One possibility
is that Co favors another magnetic order (not ferromagnetic).
Another possibility is that alloy disorder
on the Fe site induced by 25\% Co replacement is sufficiently strong to
destroy the ferromagnetism (this is neglected in our calculation, which
uses an artificial perfect ordering of Fe and Co). Related to this, it is
possible that the presence of Co-Co dimers in the alloy, but not in the
present calculation affects the behavior.
Finally, we note that Bittar and co-workers observed that
lower Co concentrations $\sim$5\% substitution bring the system
close to ferromagnetism based on the enhancement of the susceptibility.
\cite{bittar}

\section{PBE+$U$ Calculations}

As mentioned, a scenario in which semiconducting FeGa$_3$ is
antiferromagnetic with pairs of Fe atoms forming singlets
was proposed by Yin and Pickett \cite{yin}
based on LDA+$U$ calculations
and subsequently criticized by Oscorio-Guillen and co-workers, \cite{guillen}
based on theoretical considerations.
Arita and co-workers \cite{arita} used photoemission
and inverse photoemission experiments to investigate the electronic structure.
They find that the spectra can be reproduced by LDA+$U$ calculations
using values of $U$ of $\sim$3 eV. However, the density of states obtained
is very similar between such calculations and $U$=0 density functional
calculations. The data in relation to the
calculations do not clearly distinguish the LDA+$U$ and $U$=0 calculations,
and as mentioned there are some other questions that arise with an
explanation in terms of an antiferromagnetic ground state.
In any case, we did additional calculations with various values
of $U$ and different double counting schemes.

We start with calculations with the fully localized limit (SIC) double counting,
similar to what was reported by Yin and Pickett. Our PBE+$U$ calculations
were done without spin orbit.
First of all, we verified that there is no antiferromagnetic
solution without $U$. Secondly, in calculations with the SIC double
counting we do find an antiferromagnetic solution in accord with
what was reported by Yin and Pickett.
We obtain a moment in the Fe LAPW sphere (radius
2.05 Bohr) of 0.52 $\mu_B$ with $U-J$=1.4 eV,
again in accord with Yin and Pickett (who however do not state
the sphere radius over which they integrated the moment).

We also find that we can obtain
an antiferromagnetic solution with the AMF double counting, but that
substantially higher $U-J$ is needed. At $U-J$=3 eV the moment
in the LAPW sphere is still only
0.39 $\mu_B$.
Also we do not find good local moment behavior. Specifically,
the moment vanishes for imposed ferromagnetic ordering with SIC double
counting and $U-J$=1.4 eV. This implies that $U$ is stabilizing a band
structure driven antiferromagnet, which is not consistent with the 
assumptions of the fully localized limit.
We note that there is no clear justification for adding $U$ in weak or
moderately correlated transition metal compounds especially
with this double counting scheme - a point that was emphasized
by Oscorio-Guillen and co-workers. \cite{guillen} For example, Fe metal
is well described by standard density functionals, and agreement
with experiment would be degraded if one performs $+U$ calculations.

In any case, at higher $U-J$
we do obtain behavior that is closer to local moment in the sense that
both ferromagnetic and antiferromagnetic solutions can be found, but the moments
are high at that point, and clearly should have been seen in the susceptibility
if present. For $U-J$=3.0 eV,
we obtain a moment of 1.70 $\mu_B$ for the
antiferromagnetic order and 1.43 $\mu_B$ for ferromagnetic.

For the AMF double counting, while we do obtain
an antiferromagnetic solution at $U-J$=3 eV, we do not obtain a ferromagnetic
solution. This provides a resolution for the apparent discrepancy between
the the results of Yin and Pickett and those of 
Oscorio-Guillen and co-workers, who considered ferromagnetic
ordering via fixed spin moment calculations.

Thus it is possible to obtain small antiferromagnetic moments with
PBE+$U$ calculations, but these are band structure related, inconsistent
with the assumptions of the SIC double counting. At higher $U$ where local
moments are found, these are large, approaching 2 $\mu_B$, inconsistent
with experiment. We note that the use of SIC double counting LDA+$U$
calculations with low values of $U-J$ is rarely justified.
Clearly, it is not the case that there is a stable moment approximating
spin 1/2 over a range of $U$. Rather, the value used by Yin and Pickett is
a threshold value for the SIC double counting where the Fe begins a spin
state transition and its moment rapidly increases with $U$. This means
that models based on spin 1/2 Fe moments are not applicable for FeGa$_3$.
This raises doubts about whether there are such static moments
in undoped FeGa$_3$. Furthermore, as discussed above, we can obtain
ferromagnetism in doped FeGa$_3$ without assuming such pre-existing
moments. Neutron diffraction measurements for semiconducting FeGa$_3$
should be sensitive to any ordered Fe moments if present, in particular
through symmetry lowering from the full $P4_2/mnm$ spacegroup.

\section{summary and conclusions}

In summary, we presented first principles calculations of the electronic
structure and transport properties of FeGa$_3$. We find that modest
electron or hole doping can produce itinerant ferromagnetism
in this compound. This does not depend on the presence of pre-formed
moments in the undoped semiconducting phase. It will be of interest
to search experimentally for moments and to examine the nature of the
ferromagnetic phase in more detail.
Itinerant magnetism implies strong coupling between the electrons
at the Fermi energy that control transport and the magnetism.
As such, FeGa$_3$ may be a particularly interesting material near
a quantum critical point.
We find that the ferromagnetic state is half metallic over a substantial
composition range.
The results also show some promise
as a thermoelectric material at high
temperature if the doping level is optimized.

\acknowledgments

Work at ORNL was supported by the Department of Energy, Basic Energy Sciences,
Materials Sciences and Engineering Division.

\end{document}